\documentstyle[aps,preprint,tighten]{revtex}

\newcommand{\lsim}{\mathrel{\mathop{\kern 0pt \rlap
  {\raise.2ex\hbox{$<$}}}
  \lower.9ex\hbox{\kern-.190em $\sim$}}}
\newcommand{\gsim}{\mathrel{\mathop{\kern 0pt \rlap
  {\raise.2ex\hbox{$>$}}}
  \lower.9ex\hbox{\kern-.190em $\sim$}}}

\begin{document}


\preprint{
\begin{tabular}{r}
DFTT 4/98 \\
March 1998 \\
\vspace{2pc}
\end{tabular}}

\title{\Large \bf Effects of galactic dark halo rotation \\
        on WIMP direct detection}

\vspace{3pc}
\author{
\bf F. Donato$^{\mbox{\rm a}}$, N. Fornengo$^{\mbox{\rm a}}$,
S. Scopel$^{\mbox{\rm b}}$\footnote{INFN Post--doctoral Fellow}
\vspace{2pc}
}

\address{
\begin{tabular}{c}
$^{\mbox{\rm a}}$
Dipartimento di Fisica Teorica, Universit\`a di Torino \\
and INFN, Sezione di Torino, \\
via P. Giuria 1, 10125 Torino, Italy \\
{\sl donato@to.infn.it,fornengo@to.infn.it}
\\
\\
$^{\mbox{\rm b}}$ Instituto de F\'\i sica Nuclear y Altas Energ\'\i as, \\
Facultad del Ciencias, Universidad de Zaragoza, \\
Plaza de San Francisco s/n, 50009 Zaragoza, Spain \\
{\sl scopel@posta.unizar.es}
\end{tabular}
}
\maketitle

\begin{abstract}
The effects of a possible rotation of the galactic dark halo on
the calculation of the direct detection rates for particle dark matter
are analyzed, with special attention to the extraction
of the upper limits on the WIMP--nucleon scalar cross section from the
experimental data. We employ a model of dark halo rotation
which describes the maximal possible effects. For WIMP masses above
50 GeV, the upper limit exclusion plot is modified by less than a factor of two
when rotation is included. For lighter masses the effect can be stronger,
suggesting the necessity to develop specific models of halo rotation in
order to provide more accurate conclusions.
\end{abstract}

\newpage
\section{Introduction}

The possibility to detect Weakly Interacting Massive Particles (WIMPs)
distributed in the halo of our Galaxy has been a major issue
in the last years, since these particles could provide the
amount of dark matter necessary to explain many observed
dynamical properties of galaxies, clusters and of the Universe itself.
Different kinds of possible signals have been identified and looked for,
in order to outline the presence of WIMPs in our Galaxy. These signals are
usually referred to as ``direct" and ``indirect" detection rates.
Direct detection refers to the possibility to measure
a WIMP--nucleus interaction in a low--background detector,
while indirect detection relies on the measurement of
WIMP annihilation products: photons, antiprotons and positrons
produced by the annihilation in the galactic halo, or neutrinos
coming out of the Earth or the Sun where WIMPs may have been
accumulated as a consequence of gravitational capture.
It is remarkable that the present sensitivity of the different experiments
is already at the level of the predicted rates for specific WIMP candidates,
like the neutralino, which represents one of the most interesting and studied
cold relic particles \cite{noi}.

The calculation of the different detection rates depends not only on
the particle physics properties of the WIMPs interactions, but also on the
characteristics of the galactic halo where the WIMPs are distributed.
Direct detection rates and upgoing-muon fluxes
at neutrino telescopes, which both rely on the WIMP elastic scattering off
nuclei, depend on the WIMP matter density $\rho_\odot$ and
velocity distribution $f_\odot(v)$
at the Earth position $r_\odot$ in the Galaxy. In particular, the dependence
of the signals on $\rho_\odot$ is a linear one.
The other indirect signals
(photon, antiproton and positron fluxes) have a stronger dependence
on the matter distribution, since they are proportional
to the square of the matter distribution function (DF) $\rho(\vec r)$
integrated over the effective region of production and propagation
of the annihilation products. On the contrary, this kind of signals are
essentially independent on the details of the velocity DF, since
the annihilating WIMPs are almost at rest and corrections due to their
velocity dispersion are negligible.

Detailed estimates of the detection rates would require
specific and accurate models of the galactic halo able to provide
a reliable WIMPs DF $g(\vec x, \vec v)$
(not necessarily separable in phase space, i.e. $g(\vec r, \vec v) =
\rho(\vec r) f(\vec v)$). Unfortunately, detailed halo models are not
available at present, mainly because the
constraints obtained from astrophysical observations
are not stringent enough to restrict different
possibilities. The most important
observational constraint is provided by
the flatness of the rotation curves at large radii.
Although the available data on our Galaxy do not provide
a compelling evidence of a flat rotation curve, this feature is observed
in a large number of spiral galaxies and therefore it looks
reasonable to assume its validity also for our Galaxy.

The standard and simplest model of the dark galactic halo, which is
compatible with a flat rotation curve, is the so--called isothermal
sphere. This model relies on the two basic assumptions of
spherical symmetry and thermal equilibrium, which find a strong
support in the argument of ``violent relaxation" introduced by Lynden--Bell
30 years ago\cite{lynden-bell}. In this model, the 
DF is separable into a matter density distribution $\rho(r)$,
which has a $r^{-2}$ behaviour at large radii, and into a
Maxwell--Boltzmann (MB) velocity DF $f(v)$ \cite{binney}.
Although such a model gives a divergent total mass and therefore
an appropriate cut-off has to be introduced at large radii,
its range of validity has been tested at least in the inner parts
of many galactic systems. Moreover, since it
represents a simple and reasonable approximation, in the absence of a more
detailed model it is widely adopted to describe the dark halo
of our Galaxy. However, many different models are known to be consistent with
flat rotational curves.  For  instance, models  which  describe
non--spherically  symmetric or flattened halo  distributions have
been discussed  \cite{binney}. In these models, the specific form of
$\rho(\vec r)$ differs from the standard isothermal sphere matter DF,
especially at  small radii,  entailing quite large
uncertainties  on the  local value  $\rho_\odot$. A comprehensive numerical
study which takes into account a large number of models  indicates that the
local  value of the   non--baryonic dark  matter  density falls in  the (rather
conservative) range 0.1 $\lsim  \rho_\odot  \lsim$  0.7 GeV
cm$^{-3}$\cite{turner}. Contrary to the matter DF, the specific form of the
velocity DF $f(v)$ has been much less investigated.
Modifications to the standard MB velocity DF are known
\cite{binney,evans}, but the  problem of
determining  the correct form of the distribution of the WIMP velocities in the
halo has  no clear  and simple   solution at  present,  both  theoretically and
observationally.   The   velocity DF is required to be
consistent  with a given  $\rho(\vec r)$ but this,  in general,  does not
determine $f(v)$ in a unique way.

The calculation of the  WIMP detection rates is  usually performed by using
the standard  isothermal sphere model. However,
modifications in the isothermal  model  can  affect the detection rates,
introducing uncertainties in the theoretical predictions and in the extraction
of the experimental limits on the WIMPs parameters. The effects induced on the
detection rates by a modification in the 
matter DF are simple to take into account,
since the dependence of the detection rates on $\rho(\vec r)$ can be factorized.
Specifically, the physical range of $\rho_\odot$ quoted above implies an
uncertainty of about a factor of 7 
in the evaluation of the direct detection rates and in
the neutrino fluxes \cite{noi} (it has to be remarked that
this large factor reflects a rather conservative attitude).
Even larger uncertainties affect the indirect rates from
WIMP annihilation in the halo, since in this case a modification in the
matter density profile can strongly 
affect the integral of $\rho^2(\vec r)$ over the effective
production region of the signal 
\cite{pbar_to,pbar_japan,gamma}.
Contrary to the case of the matter DF, a modification of the
standard MB velocity  DF would affect
the direct detection  rates and the indirect rates at neutrino telescopes
in a much  more  involved way. 
This is because the  dependence  of these  rates on  $f(v)$ is
through a  convolution of   $f(v)$  with the  differential  WIMP--nucleus cross
section. Since the  WIMP--nucleus scattering depends on the relative velocity
of the WIMPs  with  respect to the  detector  nuclei, a potentially significant
effect could be due to a bulk rotation of the halo. This would
necessarily  modify the  WIMP phase--space  DF with
respect to the standard MB form.

In this paper we wish to discuss the possible 
effects  induced by a halo rotation on
the direct detection  rates, with special 
attention to the ensuing consequences on the
determination   of the  upper  limits  on the   WIMP--nucleus  cross  section
from the experimental data. 
A calculation of  the direct  detection rates in  the case of a  rotating halo
has been  addressed in  Ref.  \cite{kamion},  where it has  been  concluded that the
maximal effect of rotation leads to a 30\% effect on the total detection
rates for a Ge nucleus, in the case of an ideal detector with no threshold.
However,  when  considering a real  detector the  behaviour of  the differential
rates at  threshold and the detector characteristics
are crucial in  determining the  experimental  limits on the
WIMP--nucleus  cross  section \cite{noi}.  Therefore, we  explicitly take into
account the  features of running detectors,  such as thresholds, quenching
factors and energy resolution, in order to estimate the largest uncertainties 
induced by a possible halo rotation in a confident way.
To this aim, following Ref.\cite{kamion} we model the galactic rotation
as described by Lynden--Bell in Ref.\cite{lynden-bell2}, where the maximally
rotating velocity DF compatible with a given mass distribution
has been derived, on the ground of purely kinematical arguments. Even if
Lynden--Bell's model of halo rotation may not represent a situation which is
realized in a physical halo, we consider it useful to bracket the size of the
effect of halo rotation on the direct detection rates.

The plan of our paper is the following. 
In Sect.II we briefly describe the calculation
of the direct detection rates in the presence of halo rotation. In Sect.III
we discuss our results for Ge, NaI and Xe detectors,
taking into account the most recent experimental data of the different
Collaborations.
Finally, in Sect.IV we draw our conclusions. An Appendix is added, where we
report the analytical expressions of the relevant part of the direct detection
rates which contain the details
of the velocity DF in the case of the standard non--rotating,
maximally co--rotating and maximally counter--rotating haloes.

\section{Direct detection rates}

The interaction of a WIMP of mass $m_\chi$ with a detector
produces the recoil of a nucleus with energy $E_R$ of the order of
few to tens keV. The recoil energy can be measured by means of
various experimental techniques with different nuclear species.
At present, experiments are running with Ge, NaI, Xe, CaF$_2$, TeO$_2$
detectors 
\cite{ge_caldwell,ge_got,ge_heid,ge_twin,ge_cosme,nai_dama,nai_boulby,xe_dama,caf2,tellurio}
and other nuclei are currently under investigation.
The relevant quantity to be calculated and compared
with the experimental measurements is the differential detection rate
\begin{equation}
\frac {dR}{dE_R}=N_{T}\frac{\rho_{\chi}}{m_{\chi}}
                    \int \,d \vec{v}\,f(\vec v)\,v
                    \frac{d\sigma}{dE_{R}}(v,E_{R}) \label{rate}
\label{eq:diffrate0}
\end{equation}
where $N_T$ is the number of the target nuclei per unit of mass,
$\rho_\chi$ is the local WIMP matter density,
$\vec v$ and $f(\vec v)$ denote the WIMP
velocity and velocity DF in the Earth frame
($v = |\vec v|$) and $d\sigma/dE_R$ is the WIMP--nucleus differential cross section.
The nuclear recoil energy is given by
$E_R={{m_{\rm red}^2}}v^2(1-\cos \theta^*)/{m_N}$,
where $\theta^*$ is the scattering
angle in the WIMP--nucleus center--of--mass frame,
$m_N$ is the nuclear mass and $m_{\rm red}$ is the WIMP--nucleus
reduced mass.
Eq.(\ref{eq:diffrate0}) refers to the situation of a monoatomic
detector, like the Ge detectors. For more general situations, like
for instance the case of NaI, the generalization is straightforward.

The differential WIMP--nucleus cross section can be expressed as
\begin{equation}
\frac{d \sigma}{d E_R}=\frac{\sigma_0}{E_R^{\rm max}} F^2(q)
\label{eq:dsigma_dq}
\end{equation}
where $\sigma_0$ is the point-like total WIMP--nucleus cross section,
$E_R^{\rm max}$
is the maximum value of $E_R$
and $F(q)$ denotes the nuclear form factor, expressed as a function
of the momentum transfer $q^2 \equiv\mid{\vec {q}}\mid^2=2m_NE_R$.

The nuclear form factor
depends sensitively on the nature of the effective
interaction involved in the WIMP--nucleus scattering.
To be definite, in the following we will consider the case of
a WIMP--nucleus scalar interaction, since this is the one which is
currently accessible to the present sensitivity of running 
detectors\cite{noi,noi_modul}.
In this case  we use the Helm parameterization of
the scalar form factor\cite{engel}:
\begin{equation}
F(q)=3 \frac {j_1(qr_0)}  {qr_0} \exp \left(-\frac{1}{2} s^2 q^2 \right)
\label{eq:ff}
\end{equation}
\noindent
where $s \simeq 1~ \rm fm$ is the
thickness parameter for
the nucleus surface, $r_0 = (r^2-5s^2)^{1/2}$, $r=1.2~A^{1/3}$ fm,
$A$ is the nuclear mass number and
$j_1(qr_0)$ is the spherical Bessel function of index 1.

In the isothermal halo model the velocity DF is a Maxwell--Boltzmann
distribution in the galactic rest frame.
Taking into account a finite escape velocity, its
expression is the following:
\begin{equation}
f_{\rm gal}(v^{\rm gal}) = N  \left(\frac{3}{2 \pi w^{2}}\right)^{3/2}
\exp \left(- \frac{3 (v^{\rm gal})^2}{2 w^2}\right)
\label{eq:MB}
\end{equation}
where the normalization factor is
\begin{equation}
N = \left[ {\rm erf}(z)-\frac{2}{\sqrt{\pi}}
z \exp(-z^{2}) \right]^{-1}.
\label{eq:MBnorm}
\end{equation}
In the previous Eqs., $z^{2}=3v_{\rm esc}^{2}/(2w^{2})$ and
$w$ denotes the root mean square velocity.
In the isothermal sphere model, $w$ is related to the asymptotic value
$v_\infty$ of the rotational velocities by the simple relation
$w =\sqrt{\frac{3}{2}}v_\infty $.
The measured rotational velocity of the Local System at the Earth's position
is $v(r_{\odot}) = 220 \pm 20$ km sec$^{-1}$ \cite{vrot} and
remains almost flat (to roughly 15\%) between 4 Kpc and 18 Kpc.
Identifying this value
with $v_\infty$ one gets the estimate $w = 270 \pm 25$ km sec$^{-1}$.

In order to evaluate the WIMP--nucleus interaction--rate,
Eq.(\ref{eq:MB}) has to be transformed to the rest frame of the Earth,
which moves through the Galaxy with
a velocity $v_\odot = 232 \pm 20$ km sec$^{-1}$ in the azimuthal direction
(this value for $v_\odot$ takes into account the motion of the solar
system with respect to the Local System). Therefore, the velocity
$\vec v$ of the WIMPs, 
as seen in the Earth's frame, is related to their velocity in the
Galactic frame 
$\vec v^{\rm gal}$ by the following set of transformation equations:
$v_\phi = v_\phi^{\rm gal} - v_\odot$ in the azimuthal direction, and
$v_\bot = v_\bot^{\rm gal}$ and $v_R = v_R^{\rm gal}$ in the vertical ($\bot$) and
in the radial ($R$) direction along the galactic plane.

By means of the previous  definitions, the  differential rate can be written in
the  form
\begin{equation}  \frac   {dR}{dE_R}=   N_{T}
\frac{\rho_{\chi}}{m_{\chi}}\frac{m_N \sigma_0}  {2   m_{\rm   red}^2}
F^{2}(q^2)\, {\cal  I}(v_{\rm min},v_\odot,v_{\rm esc})
\label{eq:diffrate1}
\end{equation}
where the  function ${\cal  I}(v_{\rm min},v_\odot,v_{\rm esc})$ contains all
the details of the integration of the velocity DF $f(v)$
in the Earth's frame:
\begin{eqnarray}
{\cal  I}(v_{\rm min},v_\odot,v_{\rm esc})  =
\int d {\vec v} \, \frac{f(\vec{v})}{v} = 
 \int_{v_{\rm min(E_R)}}^{v_{\rm esc}} \, v\, dv\,\bar{f}(v).
\label{eq:i_func}
\end{eqnarray}
In Eq.(\ref{eq:i_func}) we have defined:
\begin{eqnarray} 
\bar{f}(v) = 2\pi \int_{(\cos\theta)_{\rm min}}^{(\cos\theta)_{\rm max}} 
d\cos\theta\, f(v,\cos\theta)
\label{eq:fbar}
\end{eqnarray}
where $(\cos \theta)_{\rm min}$ and $(\cos \theta)_{\rm max}$ depend
on $v$, $v_\odot$ and $v_{\rm esc}$. In Eq.(\ref{eq:i_func})
$v_{\rm min}(E_{R}) = (m_N E_R/(2 m^2_{\rm red}))^{1/2}$.
Moreover, we have  explicitly  considered  that  particles which
possess velocities greater than the escape velocity $v_{\rm esc}$
are not bounded to the halo. The value of $v_{\rm esc}$ is somewhat
uncertain: $v_{\rm esc} = 600 \pm 200$ Km sec$^{-1}$ \cite{vesc}. A low value of
$v_{\rm esc}$ can sizably affect the detection rates, especially
at low WIMP masses. In the next
Section, we will use, as a reference value, $v_{\rm esc} = 650$ Km sec$^{-1}$.

Eq.(\ref{eq:MB}) describes a galactic halo which does not possess a bulk
rotation. In order to analyze the effect of a possible rotation of the
isothermal sphere, we consider a class of models
discussed by Lynden--Bell \cite{lynden-bell2}
which, for any given mass distribution, describe the
fastest rotating steady state by means of the following recipe:
\begin{eqnarray}
f_{+} (v^{\rm gal}) = \left \{
\begin{array}{ll}   f(v^{\rm gal}) \hspace{15mm} & v_\phi^{\rm gal} > 0 \\
                    0              & v_\phi^{\rm gal} < 0
\end{array}
\right.
\label{eq:f_piu}
\end{eqnarray}

\begin{eqnarray}
f_{-} (v^{\rm gal}) = \left \{
\begin{array}{ll}   0     \hspace{24mm}  & v_\phi^{\rm gal} > 0\\
                    f(v^{\rm gal})       & v_\phi^{\rm gal} < 0
\end{array}
\right.
\label{eq:f_meno}
\end{eqnarray}

The LB model corresponds to an un--relaxed system with the maximal rotation
compatible with a given mass distribution.
There is no indication, neither theoretical
 nor observational, that this model can
be realized in physical galactic haloes.
 Nevertheless, since it provides the largest
rotation effect, we choose to use it in order to estimate the maximal 
modification induced by galactic rotation to the direct detection rates and to
 the extraction of the upper limits on the WIMP--nucleus cross section.

The analytic expressions for ${\cal  I}(v_{\rm min},v_\odot,v_{\rm esc})$ are
given in the Appendix for the non--rotating model of Eq.(\ref{eq:MB})
and for the models of
Eq.(\ref{eq:f_piu}) (maximal co--rotation) and Eq.(\ref{eq:f_meno})
(maximal counter--rotation). The three DFs $\bar{f}(v)$, for
the  non--rotating,  co--rotating and counter--rotating cases,  are plotted in
Fig.1. In the co--rotating situation the WIMPs have a bulk rotation in the same
azimuthal direction as the Earth. Therefore, the relative velocity
between the WIMPs and the detector is,  on average, reduced.
On  the  contrary, for a   counter--rotating  halo the  relative average
velocity increases and, moreover, there is a lower velocity cut--off
corresponding to $v_\odot$.  The reason why all the three curves in Fig.1 cross
at the same point is a feature of the particular choice of the distribution
functions in Eqs. (\ref{eq:f_piu})--(\ref{eq:f_meno}) and does not reflect any
general property of rotating models.

Eq.(\ref{eq:diffrate1}) represents the differential rate for an ideal detector.
In order to compare the calculated rates with the measured ones, we have to
express Eq.(\ref{eq:diffrate1}) as a function of the electron--equivalent
energy $E_{\rm ee}$ (which is actually measured, instead of $E_R$)\cite{smith}.
 The quantity $E_{\rm ee}$ is simply proportional to the nuclear recoil
energy  through  the  quenching factor  $Q$,  i.e. $E_{\rm ee}  = Q  E_R$.
Moreover, the energy resolution of the detector has to be taken into account.
This is obtained by means of the convolution
\begin{equation}
\frac{dR}{dE_{\rm ee}} = \frac{K}{\sqrt{2\pi} r(E_{\rm ee})
} \int_{0}^\infty \,
\exp \left(-\frac{(E_{\rm ee} - E)^2}{2 r^2(E_{\rm ee})} \right)
\frac{dR}{dE} \, dE
\end{equation}
where $K$ is the normalization factor and is given by $K=2/[1+{\rm erf}(E_{\rm
ee}/\sqrt{2}r(E_{\rm ee}))]$.
The resolution function $r(E)$ can be expressed as a function of
the energy as
\begin{equation}
\frac{r(E)}{E} = a + \frac{b}{\sqrt{E}}
\end{equation}
where the energy $E$ is expressed in keV. The constant parameters $a$ and $b$
depend on the detector and are determined experimentally.
The values of the quenching factors $Q$ and of the resolution parameters $a$
and $b$
are reported in Table 1 for the detectors considered in Sect.III. The threshold
energies $E_{\rm ee}^{\rm th}$ of the same detectors are also given in Table 1.

\section{Results}

We start our analysis by discussing the Ge detectors.
Fig.2  shows the differential rate on a  Ge detector as a  function of the
electron--equivalent energy $E_{\rm ee}$. The solid  lines refer to
WIMPs of masses  $m_\chi =  60$ GeV (upper solid line) and $m_\chi =
20$ GeV (lower solid line). For the  upper line we used, as a reference value,
$\sigma_0 = 10^{-9}$ nbarn and 
for the  lower ones $\sigma_0 = 10^{-10}$ nbarn. The
other parameters used in  the calculation of the  rates are: 
$v_\odot = 232 $ km sec$^{-1}$,
$v_{\rm esc} =$ 650 Km  sec$^{-1}$ 
and $\rho_\chi = 0.5$ GeV cm$^{-3}$.
The differential rates for a maximally co--rotating halo are
plotted as dashed lines (the upper one refers to $m_\chi = 60$ GeV,
the lower one to $m_\chi = 20$ GeV). The results for a counter--rotating
halo are reported as dotted lines.

For low energies, the
co--rotating model gives a larger rate  as compared to the non--rotating model,
since low values of $E_{\rm ee}$
mainly correspond to low WIMP velocities in the local
frame, where the co--rotating DF is  enhanced with respect to the non--rotating
model, as is shown in Fig.1.
On the contrary, for higher  values of $E_{\rm ee}$, the co--rotating rate
becomes  smaller than the   
non--rotating one, and  the  counter--rotation situation
gives the highest rates.  The difference among the  three curves increases with
the energy, the one corresponding to a co--rotating halo rapidly
diverging from the other two. We also
notice that,  for a fixed  value of  $E_{\rm ee}$, the difference between the
rotating and non--rotating situations is more pronounced for lighter WIMPs.
This  property has a
direct influence on the determination  of  the   exclusion  plot  in the
WIMP--nucleus cross section $\sigma_0$ vs. $m_\chi$ plane, since the exclusion plot
is obtained by comparing the calculated rate with the 90\% C.L.
upper limit on the counts of the detector. Therefore, because of a possible halo
rotation, we expect a larger uncertainty for lighter WIMPs in the determination of the
exclusion plots.

In Fig.2 the 90\% C.L. upper limits of two representative Ge experiments are
also plotted: the solid histogram refers to the Neuchatel experiment 
\cite{ge_got},
the dot--dashed histogram is obtained from the Twin experiment \cite{ge_twin}.
Because of the relatively fast decrease of the differential rates as a
function of $E_{\rm ee}$, the most stringent limits on the WIMP
parameters are usually obtained from the
energy bins closest to threshold energy.
For the Neuchatel experiment, the energy threshold of the detector is 1.5 keV.
In the case of Twin, the detector energy threshold is 4 keV, but the
most stringent limits are provided by the counting rates in the
energy bins above the Gallium peak (which is clearly visible
as an increase in the counting rate around $E_{\rm ee} \simeq 10$ keV). This
fact gives an effective threshold for WIMP searches of about 12 keV,
denoted by a vertical dashed line,
similar to the Ge/Heidelberg experiment \cite{ge_heid}.
Fig.2 shows that for Ge experiments with a low energy threshold
the effects of halo rotation
are less important than for the case where the threshold energy is high.
This is especially true for a counter--rotating halo and high WIMP masses.

We are now in the position of determining the upper limit on the
WIMP--nucleus cross section as a function of $m_{\chi}$
by comparing the calculated and the experimental
rates. Actually,
the direct detection technique measures the product
$(\rho_{\chi} \times \sigma_0)$
between the WIMP local density $\rho_\chi$
and the WIMP--nucleus cross section $\sigma_0$.
The single parameters  $\rho_\chi$ and $\sigma_0$ cannot be disentangled in a
direct detection measurement.
Therefore, we report our upper bounds in terms of the product $\xi\sigma$,
where the WIMP local density is parametrized as a
fraction $\xi$ of the local total dark matter
density $\rho_\odot$ ($\xi \leq 1$) and we choose as a reference
value $\rho_{\odot}$ = 0.5 GeV cm$^{-3}$.
As it was discussed in Ref.\cite{noi}, in the case of scalar coupling
it is possible to report the results as upper limits on the WIMP--nucleon cross
section (instead of a WIMP--nucleus one), which is more suitable for the
comparison among different experiments, especially
when they make use of different target nuclei.
The WIMP--nucleon scalar cross--section is defined  as \cite{noi}
\begin{equation}
\sigma^{(\rm nucleon)}_{\rm scalar} =
\left(\frac{1+m_\chi/m_N}{1+m_\chi/m_p}\right)^2
\frac{\sigma_0}{A^{2}}
\label{eq:nucleon}
\end{equation}
where $m_p$ is the proton mass.

Fig.3 shows the 90\% C.L. upper limit on the quantity
$\xi  \sigma^{(\rm  nucleon)}_{\rm scalar}$ as a
function of the WIMP mass, for the case of a non--rotating
halo  (solid   line), of   maximal   co--rotation  (dashed   line) and  maximal
counter--rotation  (dotted line). The plot is the
convolution of the most stringent limits of all the
presently available  results from Ge
detectors \cite{ge_got,ge_heid,ge_twin,ge_cosme}. 
In the WIMP mass range reported in  Fig.3, the upper limit for
masses below 25 GeV  (for   non--rotating  and    
counter--rotating  halo   models, 40  GeV for
co--rotation) is provided by the  Neuchatel experiment and for  
higher  masses the limit comes from Twin.
Fig.3 shows that the effect of rotation of the halo can strongly affect the
low mass region in the co--rotating situation. The reason for this behaviour has been
previously discussed in relation with Fig.2, where it was shown that the
detection rates for lighter WIMPs are more affected by a possible halo rotation.
On the contrary, for masses above 50 GeV the effect is contained below a factor of two,
when the  co--rotation case is  compared to the  non--rotating halo result.
The difference between the two exclusion plots is smaller for heavier WIMPs.
As reference values, the difference is of the order of 50\% for masses around
100 GeV and is reduced below 20\% for $m_\chi \gsim 300$ GeV.
For counter--rotating models,  the modification of the  exclusion plot with respect
to the  non--rotating case is always relatively small,  not  exceeding a factor of two in
the  whole  mass  range. For WIMP masses above 100 GeV, the magnitude
of the effect is very similar to the case of co--rotation.
We  have to  remind  at this   point  that  the effect
considered here refers to  situations of maximal  rotation of the galactic halo.
Plausible models of rotating haloes lie somewhere in between the two
maximal cases  discussed here, probably much closer to the non-rotating case
than to these extreme situations.
Therefore, our results  have to be considered as
maximal  possible  effects of  galactic  rotation on  the  determination of the
exclusion plots. 

Let us now discuss the case of NaI detectors.
The large mass low--background NaI detector of the DAMA/NaI Collaboration
currently provides the most stringent upper limit on the WIMP--nucleon
scalar cross section \cite{nai_dama} (except for
a narrow window around $m_\chi \simeq 15$ GeV, where the Ge detectors are
more sensitive \cite{noi}.) Fig.4 shows the differential rate for NaI
detector as a function of the electron equivalent energy $E_{\rm ee}$. The
solid line is the rate calculated for $m_\chi=60$ GeV and
$\sigma_0 = 10^{-9}$ nbarn in the case of a non rotating halo. The dashed
line refers to the co--rotating case and the dotted line to the
counter--rotating situation. The histogram is the 90\% C.L. upper limit
from the DAMA/NaI Collaboration \cite{nai_dama}. The ensuing exclusion plot
is shown in Fig.5. Also for the NaI detector, the co--rotating case deviates
significantly from the non--rotating situation for relatively low masses.
On the contrary, for counter--rotating haloes the exclusion plot remains
close to the result for a static halo. The relative
deviation $R = [\xi \sigma^{(\rm nucleon)}_{\rm scalar}]_i /
[\xi \sigma^{(\rm nucleon)}_{\rm scalar}]_{\rm non~rot}$ ($i$ stands for
co-- or counter-- rotation) between the extracted upper limits in the case of
rotation with respect to the non--rotating case, are shown in Fig.6. The dashed
(dotted) line refers to a co--rotating (counter--rotating) halo. We observe
that also in this case, for WIMP masses larger that 50 GeV the effect of
rotation affects the exclusion plot by less than a factor of 2. This effect
is actually smaller for heavier WIMPs: for $m_\chi \gsim 80$ GeV the
exclusion plot uncertainty is of the order of 20--30\%.

Finally, we show  in Fig.7 the effects  of halo rotation on
the exclusion plot  for a $^{129}$Xe  detector.
Due to the high nuclear mass, Xe detectors are in principle more sensitive
to higher WIMP masses than the NaI and Ge ones. Moreover, the quenching
factor of Xe detectors is larger
than the one of Ge and NaI detectors, and this also shifts the sensitivity
of Xe detectors to higher WIMP masses. The DAMA/Xe Collaboration
recently reported the results of the analysis on an improved statistics of
1763.2 Kg $\times$ day obtained with an enriched liquid Xe scintillator
\cite{xe_dama}. The exclusion 
plot obtained from the data of Ref.\cite{xe_dama} is plotted in Fig.7, for the
conservative value of Q=0.44. Similar to the cases
previously discussed, for a counter--rotating halo the
deviation is always smaller than a factor of two. The same situation
happens for co--rotating models and WIMP masses larger than about 50 GeV.
Regardless of the model of rotation, for $m_\chi \gsim 200$ GeV the
deviation is always smaller than 25\%.

\section{Conclusions}

In this paper we have investigated the effect induced by a possible
rotation of the galactic halo on the rates of WIMP direct detection.
 In particular, we have discussed
the implication of halo rotation on the determination of the exclusion plots on
the WIMP--nucleon cross section for different detectors, namely
Ge, NaI and Xe ones.
The rotation of the halo has been described by using
a model \cite{lynden-bell2} which corresponds
to a situation where the halo
possesses the maximal rotation compatible with a given mass
DF, which for simplicity we have chosen to be that of
the isothermal sphere.

We found that the exclusion plots  obtained from the data are affected by  less
than a factor of 2 in the  case of  counter--rotating models. The same  size of
uncertainty  occurs also for  the co--rotating  models, when  the WIMP  mass is
larger than  about 50  GeV. For  lighter WIMPs and  co--rotation, the exclusion
plots are  modified  by a larger  amount. We  have to  remind  that, due to the
particular model of halo  rotation which we have  employed here,  these are
expected
to be maximal  effects. For  specific physical  rotation  models, the effect of
halo rotation  will be  plausibly smaller. We  can therefore  conclude that, at
least for  WIMP  masses greater  than about  50 GeV,  the  determination of the
exclusion  plots from  the  experimental  data are  affected by an  uncertainty
smaller  than  a factor of two  due to  the possibility
  that the  galactic halo
rotates,  independently on the  specific   
model of halo  rotation. 
We notice that recent preliminary data from accelerators indicate that
the lower limit on the mass of 
the most plausible WIMP candidate, the neutralino, is $m_\chi\simeq 30$ GeV 
for low value of the susy parameter $\tan\beta$, and
$m_\chi\simeq 45$ GeV for $\tan\beta\gsim 3$ \cite{lep183}.
Therefore,
for this dark matter candidate, the uncertainty on the exclusion plot
due to a possible rotation of the halo is expected to be relatively small.
The situation  is different for lighter WIMPs.  
In this  case,  it would  be required to
develop  specific models  of halo  rotation in  order to obtain  more accurate
conclusions.

\section*{Acknowledgements}

We wish to thank Sandro Bottino for many useful
discussions and encouragement in the preparation of this paper. We
also like to thank Rita Bernabei for illuminating discussions
about the direct detection technique.

\appendix
\section{}

In this Appendix we report the analytical expression of the function
${\cal  I}(v_{\rm min},v_\odot,v_{\rm esc})$ which enters in the calculation of
the differential rate for direct detection Eq.(\ref{eq:diffrate1}),
for a velocity DF which is Maxwellian in the galactic frame
(see Eq.(\ref{eq:MB})). The cases of a non--rotating, maximally co--rotating
and maximally counter--rotating halo models are given.
In all the following expressions
we make use of the dimensionless variables:
$x^{2} = 3v^{2}/(2w^{2})$, $\eta^{2}$ $=$
$  3v_{\odot}^{2}/(2w^{2})$, $z^{2} = 3v_{\rm esc}^{2}/(2w^{2})$,
$x_{\rm min} = (3 m E_{R}/4m_{\rm red}^{2}w^{2})^{1/2}$, where
$v_{\odot}$ denotes the Earth velocity in the galactic frame,
$v_{\rm esc}$ is the escape velocity,
$w$ is the root mean square velocity of the Maxwellian distribution in the
galactic rest frame. In the definition of the three distributions, the same
value of $w$ has to be used, since they correspond to the same matter density
and therefore they contribute in the same way to the rotational velocities.
We also define the function
\begin{equation}
\chi(x,y)=\frac{\sqrt{\pi}}{2} [{\rm erf}(y)-{\rm erf}(x)]
\end{equation}
where
\begin{equation}
{\rm erf}(x)=\frac{2}{\sqrt{\pi}}\int^{x}_{0}\,\exp(-t^{2})\,dt\,.
\end{equation}
The normalization constant $N$ of the Maxwellian DF is given in Eq.(\ref{eq:MBnorm}).

\subsection{Non rotating model}
\label{app:non}

\begin{eqnarray}
&& {\cal  I}(v_{\rm min},v_\odot,v_{\rm esc})=
\frac{N}{\eta} \left(\frac{3}{2\pi w^2}\right)^{1/2} \times \nonumber \\
&&\left\{
\begin{array}{lll}
\chi(x_{\rm min}-\eta,x_{\rm min}+\eta)-2\eta\,\exp (-z^{2})
                                      &\hspace{5pt}& x_{\rm min}<z-\eta  \\
\chi(x_{\rm min}-\eta,z)- \exp (-z^{2}) (z+\eta-x_{\rm min})
                                      &\hspace{5pt}&z-\eta\leq x_{\rm min}<z+\eta \\

0                                     &\hspace{5pt}&  x_{\rm min}\geq z+\eta
\end{array}
\right.
\end{eqnarray}

\subsection{Maximally co--rotating Lynden--Bell model}

In this case there are two possible situations, depending on the relative
magnitude of $v_\odot$ and $v_{\rm esc}$:

\begin{displaymath}
{\mathbf i)\,\, 2\,\eta \le z}
\end{displaymath}

\begin{eqnarray}
&&{\cal  I}(v_{\rm min},v_\odot,v_{\rm esc})=
2 \frac{N}{\eta} \left(\frac{3}{2\pi w^2}\right)^{1/2} \times \nonumber \\
&&\left\{
\begin{array}{lll}
\chi(x_{\rm min}-\eta,x_{\rm min}+\eta)+\chi(z,0)+ \exp (\eta^{2})
\chi(\eta,\sqrt{z^{2}+\eta^{2}})\\
\hspace{25pt} - \exp (-z^{2}) (\eta-z+\sqrt{z^{2}+\eta^{2}})            &\hspace{5pt}& x_{\rm min}\le\eta\\

\chi(z,x_{\rm min}+\eta)+
\exp (\eta^{2}) \chi(x_{\rm min},\sqrt{z^{2}+\eta^{2}})\\
\hspace{25pt} - \exp (-z^{2}) (\eta-z+\sqrt{z^{2}+\eta^{2}})
                                                 &\hspace{5pt}& \eta < x_{\rm min}\le z-\eta \\

\exp (\eta^{2}) \chi(x_{\rm min},\sqrt{z^{2}+\eta^{2}})
 - \exp (-z^{2}) (\sqrt{z^{2}+\eta^{2}}-x_{\rm min})
                                    &\hspace{5pt}& z-\eta<x_{\rm min}\le\sqrt{z^{2}+\eta^{2}} \\

0                                   &\hspace{5pt}&  x_{\rm min}\geq \sqrt{z^{2}+\eta^{2}}
\end{array}
\right.
\end{eqnarray}

\begin{displaymath}
{\mathbf ii)\,\, 2\,\eta \ge z}
\end{displaymath}

\begin{eqnarray}
&&{\cal  I}(v_{\rm min},v_\odot,v_{\rm esc})=
2 \frac{N}{\eta} \left(\frac{3}{2\pi w^2}\right)^{1/2}\times \nonumber \\
&&\left\{
\begin{array}{lll}
\chi(x_{\rm min}-\eta,x_{\rm min}+\eta)+\chi(z,0)+ \exp (\eta^{2})
\chi(\eta,\sqrt{z^{2}+\eta^{2}})\\
\hspace{25pt} - \exp (-z^{2}) (\eta-z+\sqrt{z^{2}+\eta^{2}})
                                                &\hspace{5pt}& x_{\rm min}\le z-\eta\\

\chi(x_{\rm min}-\eta,0) + \exp (\eta^{2}) \chi(\eta,\sqrt{z^{2}+\eta^{2}})\\
\hspace{25pt} - \exp (-z^{2}) (\sqrt{z^{2}+\eta^{2}}-x_{\rm min})
                                                &\hspace{5pt}& z-\eta < x_{\rm min}\le \eta \\

\exp (\eta^{2}) \chi(x_{\rm min},\sqrt{z^{2}+\eta^{2}})
-\exp (-z^{2}) (\sqrt{z^{2}+\eta^{2}}-x_{\rm min})
                                 &\hspace{5pt}& \eta < x_{\rm min}\le \sqrt{z^{2}+\eta^{2}} \\

0                         &\hspace{5pt}&  x_{\rm min}\geq \sqrt{z^{2}+\eta^{2}}

\end{array}
\right.
\end{eqnarray}

\subsection{Maximally counter--rotating Lynden--Bell model}

\begin{eqnarray}
&&{\cal  I}(v_{\rm min},v_\odot,v_{\rm esc})=
2 \frac{N}{\eta} \left(\frac{3}{2\pi w^2}\right)^{1/2}\times \nonumber \\
&&\left \{
\begin{array}{lll}
\chi(0,z)+ \exp (\eta^{2}) \chi(\sqrt{z^{2}+\eta^{2}},\eta)\\
\hspace{25pt} - \exp (-z^{2}) (\eta+z-\sqrt{z^{2}+\eta^{2}})
                                         &\hspace{5pt}& x_{\rm min}\le \eta\\
\chi(x_{\rm min}-\eta,z)+\exp (\eta^{2}) \chi(\sqrt{z^{2}+\eta^{2}},x_{\rm min})\\
\hspace{25pt} - \exp (-z^{2}) (\eta+z-\sqrt{z^{2}+\eta^{2}})
                                  &\hspace{5pt}& \eta < x_{\rm min}\le \sqrt{z^{2}+\eta^{2}} \\

\chi(x_{\rm min}-\eta,z)- \exp (-z^{2}) (z+\eta-x_{\rm min})
                               &\hspace{5pt}& \sqrt{z^{2}+\eta^{2}} < x_{\rm min}\le z+\eta \\
0                         &\hspace{5pt}&  x_{\rm min}\geq z+\eta
\end{array}
\right.
\end{eqnarray}

\newpage

\begin{table}
\caption{Threshold energy $E_{\rm ee}^{\rm th}$, quenching factor $Q$ and resolution parameters $a$
and $b$ for the detectors considered in the text.}
\vspace{10mm}
\begin{tabular}{|c|c|c|c|}
 Detector & $E_{\rm ee}^{\rm th}$ (keV) & $Q$ & Resolution parameters  \\
\hline \hline
DAMA/NaI
&
2
&
  \begin{tabular}{lll}
    Na  &: &0.31 \\
    I   &: &0.09 \\
  \end{tabular}
&
  \begin{tabular}{lll}
    $a$  &= &0 \\
    $b$  &= &0.579 \\
  \end{tabular}
\\
\hline
Ge/Neuchatel
&
1.5
&
0.25
&
  \begin{tabular}{lll}
    $a$  &= &0 \\
    $b$  &= &0.17~ \\
  \end{tabular}
\\
\hline
Ge/Twin
&
4
&
0.25
&
  \begin{tabular}{lll}
    $a$  &= &0 \\
    $b$  &= &0.05~ \\
  \end{tabular}
\\
\hline
DAMA/Xe
&
13
&
0.44
&
  \begin{tabular}{lll}
    $a$  &= &0.056 \\
    $b$  &= &1.191 \\
  \end{tabular}
\end{tabular}
\end{table}

\newpage

\begin{center}
\begin{large}
FIGURE CAPTIONS
\end{large}
\vspace{5mm}
\end{center}

\begin{itemize}

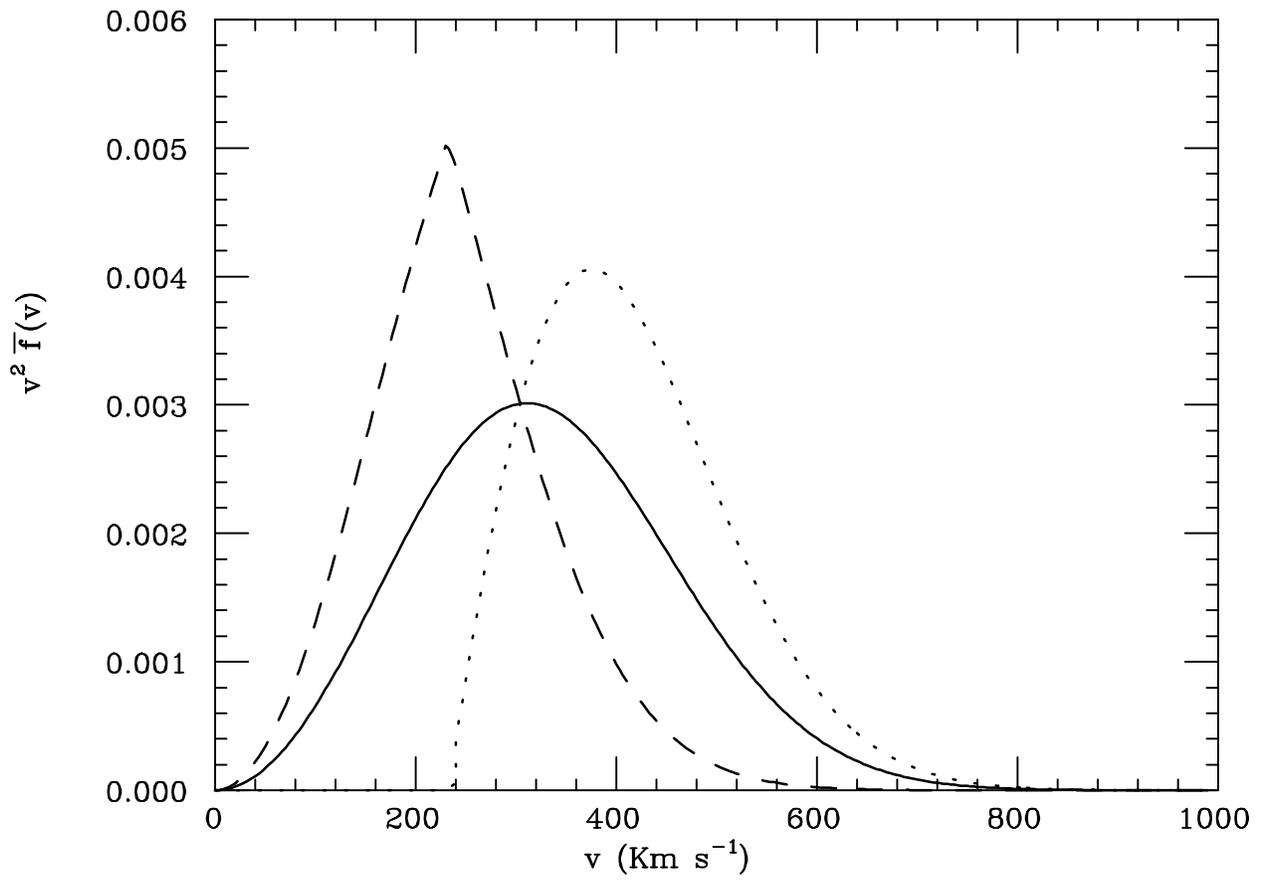
\item [FIG. 1.]
Maxwell--Boltzmann velocity DF in the local frame
(arbitrary units).
Solid, dashed and dotted lines refer respectively to the non-rotating,
maximal co--rotating and maximal counter--rotating case.

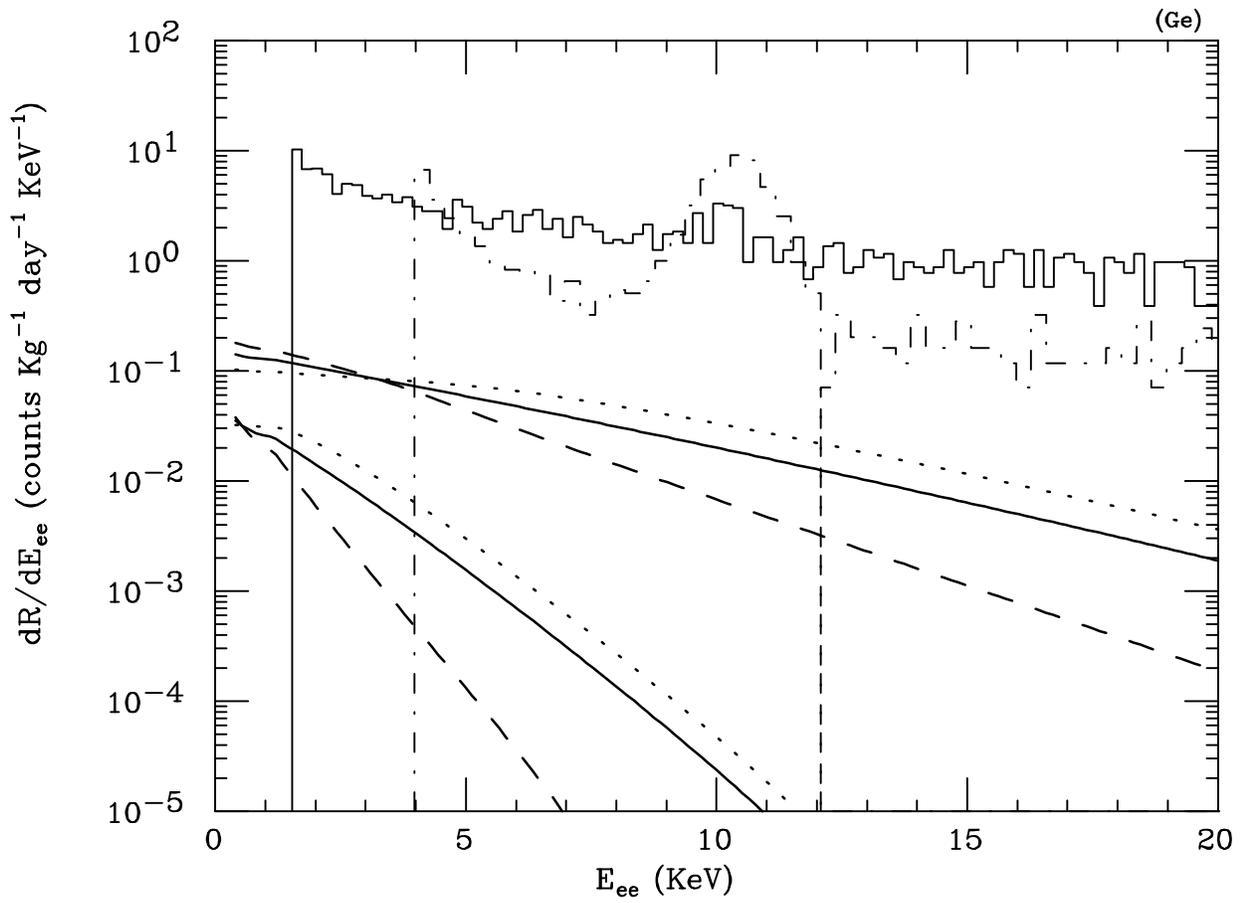
\item [FIG. 2.]
Differential rate on a Ge detector as a function of the electron--equivalent
energy $E_{\rm ee}$. The solid, dashed and dotted lines refer respectively to the
non-rotating, co--rotating and counter--rotating case.
The upper lines refer to $m_\chi = 60$ GeV and $\sigma_0 = 10^{-9}$ nbarn
The lower curves are for $m_\chi = 20$ GeV and $\sigma_0 = 10^{-10}$ nbarn.
The solid (dot--dashed) histogram represents the 90\% C.L. upper limit
on the counting rate of the Neuchatel \cite{ge_got} (Twin \cite{ge_twin})
experiment.

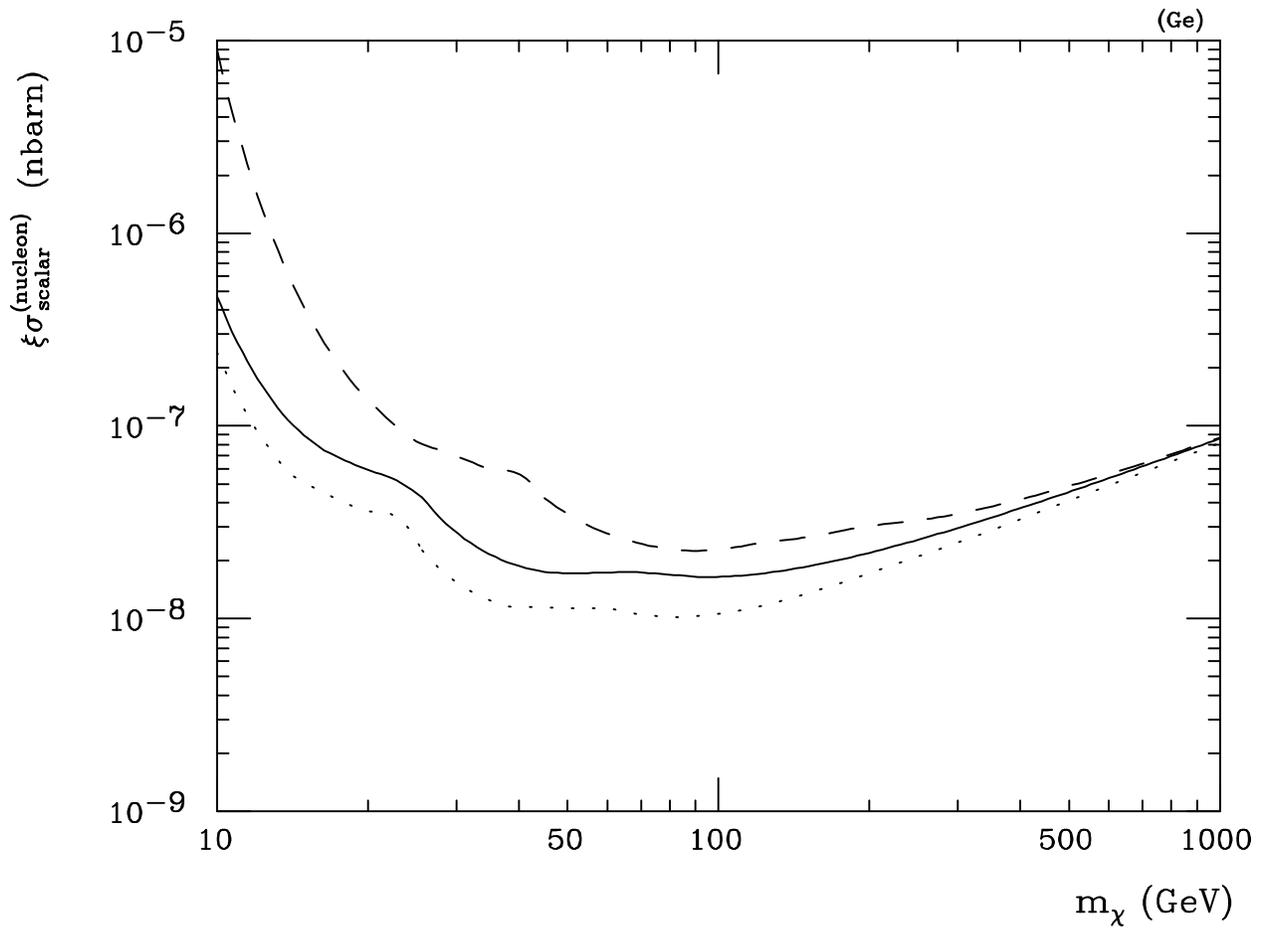
\item [FIG. 3.]
The 90\% C.L. upper limit on the quantity 
$\xi\sigma^{(\rm  nucleon)}_{\rm scalar}$
as a function of the WIMP mass $m_\chi$ for the Ge detectors 
\cite{ge_got,ge_heid,ge_twin,ge_cosme}. 
The solid, dashed and dotted lines refer to a non--rotating, co--rotating and
counter--rotating halo, respectively.

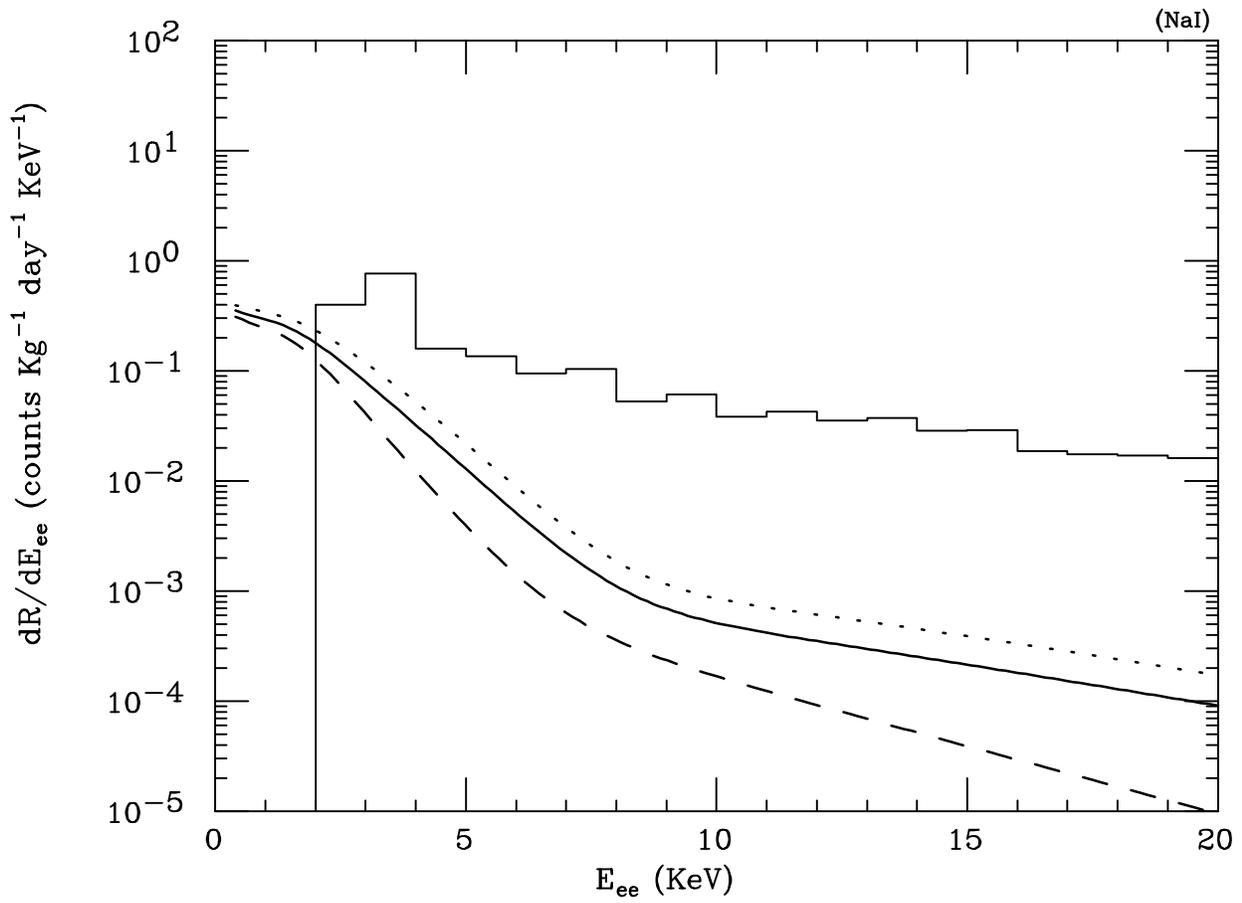
\item [FIG. 4.]
Differential rate on a NaI detector as a function of the electron equivalent
energy $E_{\rm ee}$, for a WIMP of $m_\chi=60$ GeV and $\sigma_0 = 10^{-9}$ nbarn.
The solid, dashed and dotted lines refer  to the
non-rotating, co--rotating and counter--rotating case, respectively. 
The histogram represents the 90\% C.L. upper limit on the
counting rate obtained from the DAMA/NaI
Collaboration \cite{nai_dama}.

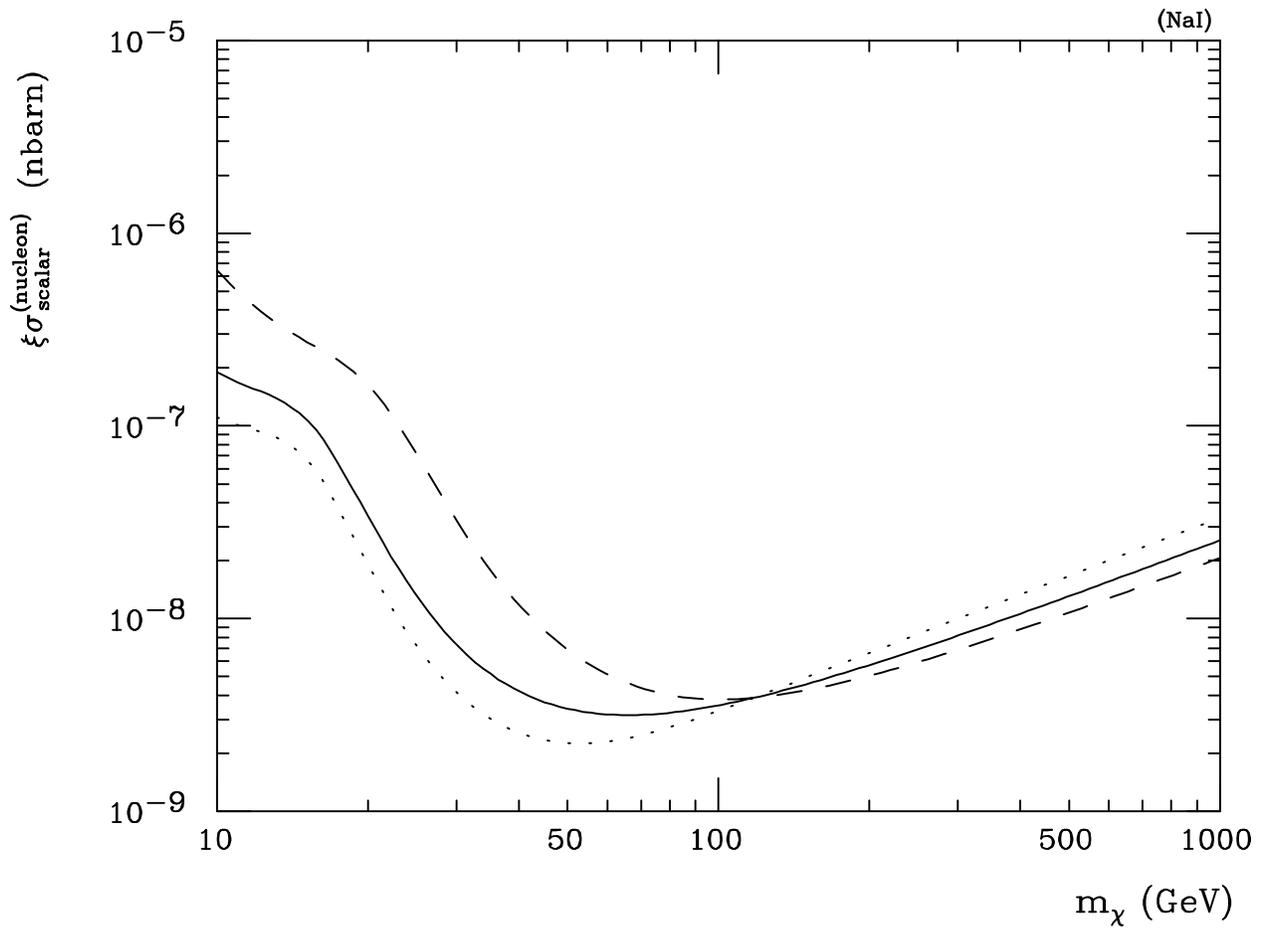
\item [FIG. 5.]
The 90\% C.L. upper limit on the quantity $\xi\sigma^{(\rm  nucleon)}_{\rm scalar}$
as a function of the WIMP mass $m_\chi$ for the DAMA/NaI detector
\cite{nai_dama}.
The solid, dashed and dotted lines refer to a non--rotating, co--rotating and
counter--rotating halo, respectively.

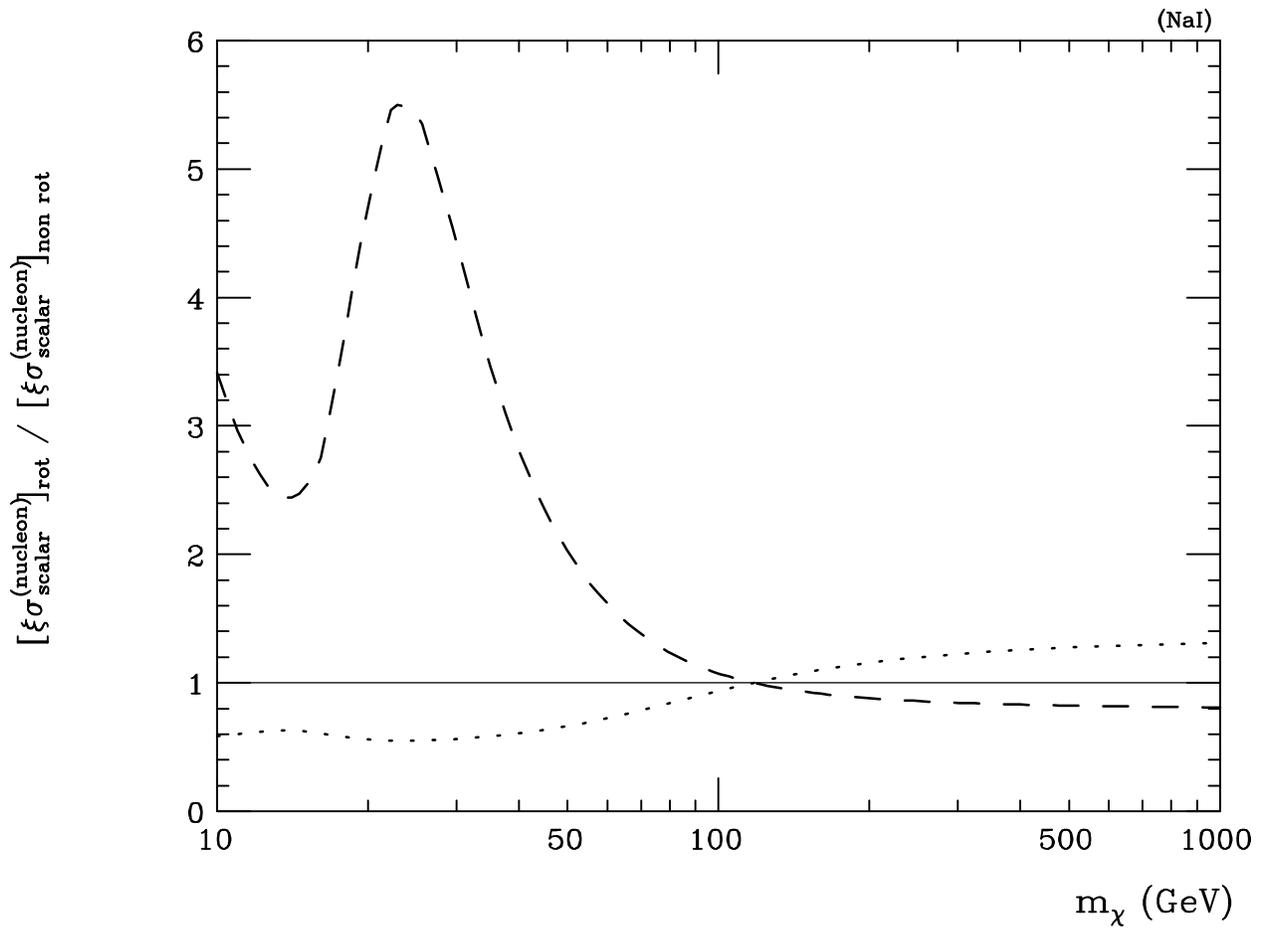
\item [FIG. 6.]
Relative deviation $R = [\xi \sigma^{(\rm nucleon)}_{\rm scalar}]_{\rm rot} /
[\xi \sigma^{(\rm nucleon)}_{\rm scalar}]_{\rm non~rot}$
between the 90\% C.L. upper limits in the case of a
rotating and a non--rotating halo, for the DAMA/NaI detector.
The dashed (dotted) line refers to the co--rotating (counter--rotating)
halo model.

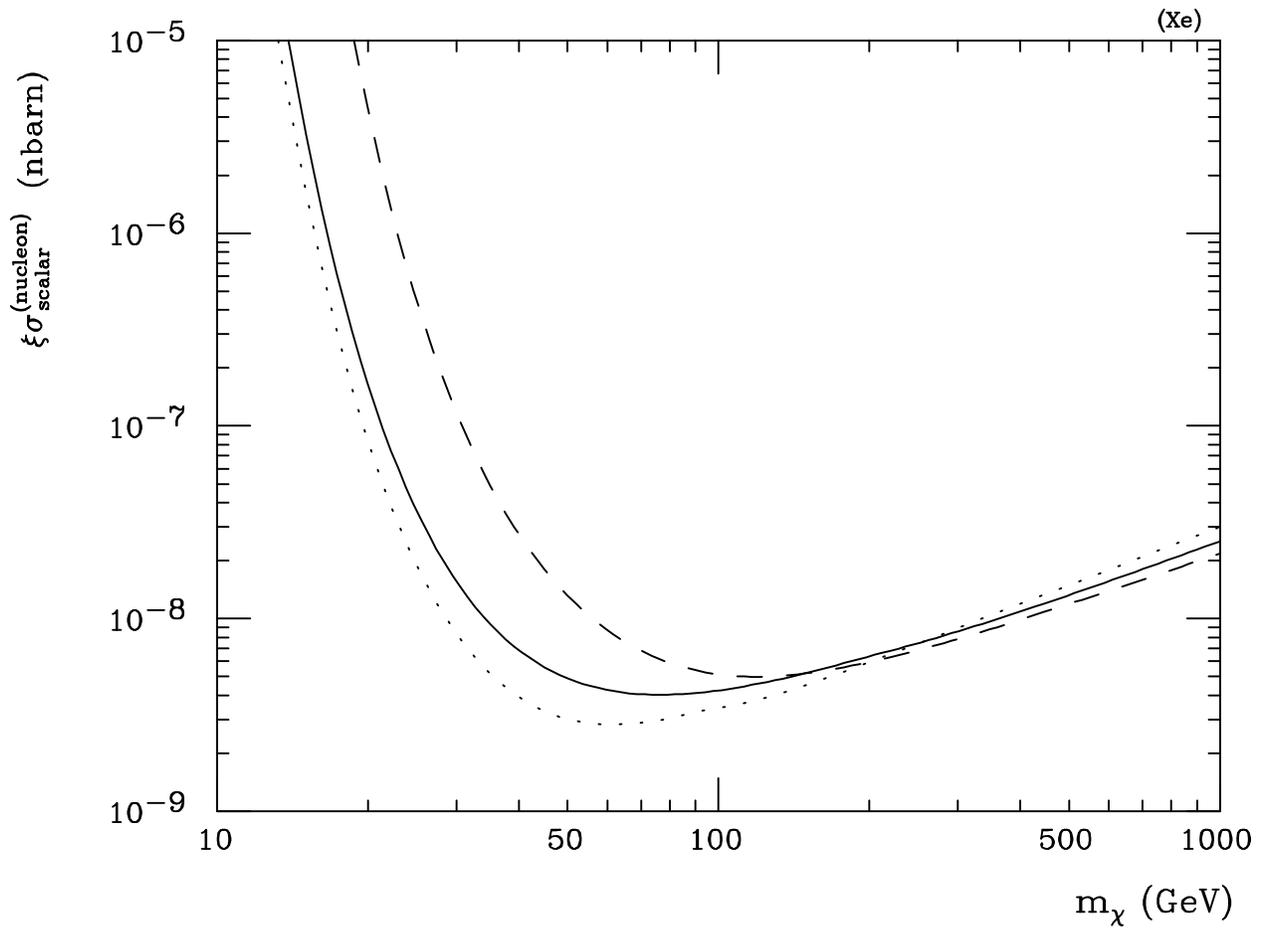
\item [FIG. 7.]
The 90\% C.L. upper limit on the quantity $\xi\sigma^{(\rm  nucleon)}_{\rm scalar}$
as a function of the WIMP mass $m_\chi$ for the DAMA/Xe detector \cite{xe_dama}.
The solid, dashed and dotted lines refer to a non--rotating, co--rotating and
counter--rotating halo, respectively.

\end{itemize}


\begin{thebibliography}{99}

\bibitem{noi} A. Bottino, F. Donato, G. Mignola, S. Scopel, P. Belli and A.
              Incicchitti, {\it Phys. Lett.} {\bf B} {\bf 402} (1997) 113;
              N. Fornengo,
              Proc. of the Int. Workshop ``Physics Beyond the Standard Model:
              from theory to experiment (Valencia97)'', Valencia, October 1997
              and Proc. of the Workshop ``DM--Italia 97: Dark matter:
              perspectives and projects", Trieste, December 1997.

\bibitem{lynden-bell} D. Lynden--Bell, {\it MNRAS} {\bf 136} (1967) 101.


\bibitem{binney}    J. Binney and S. Tremaine, {\it Galactic Dynamics},
                    Princeton University Press, Princeton, 1987.

\bibitem{turner} E. I. Gates, G. Gyuk and M. S. Turner,
                    {\it Ap.J} {\bf 449} (1995) L123.

\bibitem{evans} N. W. Evans, {\it MNRAS} {\bf 260} (1993) 191.

\bibitem{pbar_to}   A. Bottino, C. Favero, N. Fornengo and G. Mignola ,
                    {\it Astrop. Phys.} {\bf 7} (1995) 73.

\bibitem{pbar_japan} T. Mitsui, K. Maki, S. Orito, {Phys. Lett.} {\bf  B389} 
                    (1996) 169.

\bibitem{gamma} L. Bergstr\"om, P. Ullio, J. H. Buckley {\tt astro-ph/9712318},
                     December 1997.

\bibitem{kamion} M. Kamionkowski and A. Kinkhabwala, {\tt hep-ph/9710337}, 
                      October 1997.

\bibitem{lynden-bell2} D. Lynden--Bell, {\it MNRAS} {\bf 120} (1960) 204.

\bibitem{ge_caldwell} S. P. Ahlen et al., {\it Phys. Lett.} {\bf B195} (1987)
603; D. O. Caldwell et al., {\it Phys. Rev. Lett.} {\bf 61} (1988) 510.

\bibitem{ge_got} D. Reusser et al., {\it Phys. Lett.} {\bf B255} (1991) 143.

\bibitem{ge_heid} M. Beck (Heidelberg--Moscow  Collaboration),
{\it Nucl. Phys.}
{\bf B} (Proc. Suppl.) {\bf35} (1994) 150; M. Beck et al., {\it Phys. Lett.}
 {\bf B336} (1994) 141.

\bibitem{ge_twin} A.K. Drukier et al.,
{\it Nucl. Phys.} {\bf B} (Proc. Suppl.) {\bf 28A} (1992) 293; 
I. R. Sagdev, \mbox{A. K. Drukier}, D. J. Welsh, A. A. Klimenko,
S. B. Osetrov, A. A. Smolnikov,
{\it Nucl. Phys.} {\bf B} (Proc. Suppl.) {\bf 35} (1994) 175.

\bibitem{ge_cosme} E. Garcia et al., {\it Nucl. Phys.} {\bf B} 
(Proc. Suppl.) {\bf 28A} (1992) 286; M. L. Sarsa et al.,
{\it Nucl. Phys.} {\bf B} (Proc. Suppl.) {\bf 35} (1994) 154;
E. Garcia et al., Proc. of  ``The Dark Side of the Universe", Rome 1994,
eds. R. Bernabei and C. Tao (World Scientific, Singapore 1994), p. 216.

\bibitem{nai_dama} R. Bernabei et al., {\it Phys. Lett.} {\bf B389} (1996) 757.

\bibitem{nai_boulby} P.F. Smith et al., {\em Phys. Lett.} {\bf B379} (1996) 299.

\bibitem{xe_dama} R. Bernabei et al., ROM2F/98/08, February 1998.

\bibitem{caf2} R. Bernabei et al., {\it Astrop. Phys.} {\bf 7} (1997) 73.

\bibitem{tellurio} A. Alessandrello et al., {\em Phys. Lett.} {\bf B384}
                    (1996) 316.

\bibitem{noi_modul} A. Bottino, F. Donato, N. Fornengo, S. Scopel, DFTT 49/97
                   {\tt hep-ph/9709292}, to appear in {\it Phys. Lett.} {\bf B}.

\bibitem{engel}    R. H. Helm, {\it Phys. Rev.} {\bf D 104} (1956) 1466; 
                    J. Engel, {\it Phys. Lett.} {\bf B264} (1991) 114.

\bibitem{vrot}   F. J. Kerr and D. Lynden--Bell, {\it MNRAS} 
                    {\bf 221} (1986) 1023.

\bibitem{vesc} P. J. T. Leonard and S. Tremaine, {\it Ap. J.}
                    {\bf 353} (1990) 486.

\bibitem{smith}  P. F. Smith and J. D. Lewin, {\it Phys. Rep.} 
                    {\bf 187} (1990) 203.

\bibitem{lambda_max}  J. Barnes and G. Efstathiou, {\it Ap. J.} 
                     {\bf 319} (1987) 575;
                      M.S. Warren, P.J. Quinn, J.K. Salmon and W. H. Zurek,
                     {\it Ap. J.} {\bf 399} (1992) 405.
 
                
\bibitem{lep183} P. Dornan (ALEPH Collaboration), presented at LEPC Conference,
                 November 1997;
                 P. Charpentier (DELPHI Collaboration), ibid.;
                 M. Pohl (L3 Collaboration), ibid.;
                 A. Honma (OPAL Collaboration), ibid. 

\end{thebibliography}
\end{document}